\documentclass{iopart}
\usepackage{iopams,psfig,graphicx}
\def\<{\langle} \def\>{\rangle}
\begin{document}
\title{Evolution of twin-beam in active optical media}
\author{Matteo G A Paris} \address{Quantum Optics $\&$ Information Group, 
INFM Udr Pavia, Italy}
\begin{abstract}
We study the evolution of twin-beam propagating inside active media 
that may be used to establish a continuous variable entangled channel 
between two distant users. In particular, we analyze how entanglement 
is degraded during propagation, and determine a threshold value for 
the interaction time, above which the state become separable, and thus 
useless for entanglement based manipulations. We explicitly calculate the 
fidelity for coherent state teleportation and show that it is larger 
than one half for the whole range of parameters preserving entanglemenent.
\end{abstract}
The crucial ingredient of quantum information is entanglement, 
which is the essential resource for quantum computing, 
teleportation, and cryptographic protocols. Entanglement is known to 
to be a valuable resource for improving measurements \cite{entame}, 
spectroscopy \cite{spectr,fabre}, lithography \cite{qlit}, interferometry 
\cite{entdec}, and tomography of quantum devices \cite{tomochannel}. 
Most of these concepts were initially developed for discrete quantum 
variables, in particular quantum bits. Recently, however, much 
attention has been devoted to the use of continuous variables (CV), 
especially Gaussian state of light by means of linear 
optical circuits \cite{koro}, since CV may be easier to manipulate than quantum bits 
in order to perform various information processes \cite{polz}. 
In the optical implementation of quantum information 
processing, the most relevant entangled channel is provided by the
the twin-beam state (also called two-mode squeezed vacuum) 
of two modes of radiation, which, in the Fock bases, is expressed as 
\begin{eqnarray}
|X\>\>=\sqrt{1-x^2}\sum_p\: x^p\: |p\>\otimes|p\>
\label{twb}\;.
\end{eqnarray}
The twin-beam is the maximally entangled state (for a given, 
finite, value of energy) of two modes of radiation. It can be produced 
either by mixing two single-mode squeezed vacuum (with orthogonal squeezing 
phases) in balanced beam splitter or, from the vacuum, by a 
nondegenerate parametric optical amplifier (NOPA). The 
evolution operator of the NOPA reads as follows 
$U_\lambda = \exp{\left[\lambda \left(a^\dag b^\dag-ab\right)\right]}$ 
where the "gain" $\lambda$ is proportional to the interaction-time, 
the nonlinear susceptibility, and the pump intensity. 
We have $x=\tanh \lambda$, whereas the number of photons of the 
twin-beam is $N=2\sinh^2 \lambda=2x^2/(1-x^2)$. In view of the duality 
squeezing/entanglement via balanced beam-splitter the parameter $\lambda$ 
is sometimes referred to as the squeezing parameter of the twin-beam (\ref{twb}).
\par
The Wigner function $W_0(x_1,y_1;x_2,y_2)$ of a twin-beam is Gaussian, 
and is given by (in the following we will omit the arguments, and 
denote the Wigner function by $W$)
$$
W_0=\left(2\pi \sigma_+^2 \: 
2\pi \sigma_-^2\right)^{-1}\:
\exp\left[-\frac{(x_1+x_2)^2}{4\sigma_+^2}
-\frac{(y_1+y_2)^2}{4\sigma_-^2} -\frac{(x_1-x_2)^2}{4\sigma_-^2}
-\frac{(y_1-y_2)^2}{4\sigma_+^2}\right]
$$
where the variances are given by  
\begin{eqnarray}
\sigma^2_+=1/4\exp\{2\lambda\} \qquad 
\sigma^2_-=1/4\exp\{-2\lambda\}
\label{stwb}\;.
\end{eqnarray}
In applications such teleportation or cryptography one needs to transfer
entanglement among distant partners, and therefore to transmit
entangled states along some kind of channel. For optical implementation 
this is usually accomplished by means of (active) optical fibers. As a 
matter of fact, the propagation of twin-beam in optical media 
unavoidably lead to degradation of entanglement due to decoherence 
induced by losses and noise. In this paper, we study the evolution of
twin-beam in active optical media, such the pair of optical fibers
that may be used to transmit twin-beam, and analyze the separability 
of the evolved state as a function of the fiber parameters. 
A threshold value for the interaction time, above which the 
entanglement is destroyed, will be analytically derived.  
\par 
In \cite{hiroshima} and \cite{mista} the decoherence of twin-beam due to 
amplitude damping  has been studied 
numerically in terms of the relative entropy of entanglement and the 
Bell nonlocality factor respectively. 
Here we analyze the propagation of twin-beam, in both active and passive
fibers, using the Wigner function.  In this way we are able to analytically 
study the evolution as  well as the entanglement properties. In particular, 
we are able to check the positivity of the partial transpose (PPT), which 
is a necessary and sufficient condition for separability for two-mode 
Gaussian state of light. For some special choices of the parameters our results 
may be compared, and are in agreement, to that of Ref. 
\cite{jeo} where the nonlocality Bell factor is computed from the Wigner 
function, and of Ref. \cite{wel} where the PPT condition is applied to the 
transformation induced by a generic four-port (amplifying or absorbing) 
optical device. 
\par 
In the following we first study the propagation of twin-beam in active
media by transforming the corresponding Master equation into a Fokker-Planck
equation for the two-mode Wigner function. Then, we apply
the PPT condition to check separability and determine a threshold value
$t_s$ for the interaction time, above which the state is separable and thus
useless for the purposes of quantum information processing. 
\par 
The propagation of a twin-beam inside active media
can be modeled as the coupling of each part of the 
twin-beam with a non zero temperature reservoir. 
The dynamics can be described in terms of the two-mode 
Master equation
\fl \begin{eqnarray} \fl
\frac{d\varrho_t}{dt} \equiv {\cal L} \varrho_t = 
\Gamma_a (1+M_a) L[a] \varrho_t + \Gamma_b (1+M_b) L[b] \varrho_t  + 
\Gamma_a M_a L[a^\dag ] \varrho_t + \Gamma_b  M_b  L[b^\dag] \varrho_t  
\label{master} \end{eqnarray}
where $\varrho_t\equiv\varrho (t)$, $\Gamma_a=\Gamma_b=\Gamma$ 
denotes the (equal) damping rate, $M_a=M_b=M$ the number of 
background thermal photons, and $L[ O ]$ is the Lindblad superoperator 
$$L[ O ] \varrho_t =  O \varrho_t  O^\dag - \frac{1}{2} O^\dag  O 
\varrho_t - \frac{1}{2} \varrho_t O O^\dag\:.$$
The terms proportional to $L[a]$ and  $L[b]$ describe the losses, 
whereas the terms proportional to $L[a^\dag]$ and $L[b^\dag]$ describe 
a linear phase-insensitive amplification process. 
This can be due either to fiber dynamics or to thermal 
hopping; in both cases no phase information is carried.
Of course, the dynamics inside the two fibers are independent on each other. 
\par
The master equation (\ref{master}) can be transformed into a Fokker-Planck
equation for the two-mode Wigner function $W (x_1,y_1;x_2,y_2)$ 
Using the differential representation of the superoperators in Eq. 
(\ref{master}) the corresponding Fokker-Planck equation reads as follows 
\begin{eqnarray}
\fl\partial_\tau W_\tau = \left[ \frac{1}{8}%
\left(\sum_{j=1}^2\partial^2_{x_j x_j} + \partial^2_{y_j y_j}\right) 
+ \frac{\gamma}2 \left(\sum_{j=1}^2 \partial_{x_j} x_j+\partial_{y_j}  y_j + 
\right) \right] W_\tau \label{fp}\:,  
\end{eqnarray}
where $\tau$ denotes the rescaled time $\tau=\Gamma/\gamma\:t$, and the 
drift term $\gamma$ is given by
\begin{eqnarray}
\gamma= \frac{1}{2M+1}\label{gamma}\:.
\end{eqnarray}
The solution of Eq. (\ref{fp}) can be written as 
\begin{eqnarray}\fl
W_\tau &=& \int _{}dx^{\prime}_1\int
_{}dx^{\prime}_2 \int _{}dy^{\prime}_1\int _{}dy^{\prime}_2 \;\:
W_0(x^{\prime}_1,y^{\prime}_1;x^{\prime}_2,y^{\prime}_2)\: 
\prod_{j=1}^2 G_\tau(x_j|x^{\prime}_j) 
G_\tau(y_j|y^{\prime}_j) \: 
\label{conv}
\end{eqnarray}
where $W_0(x_1,y_1;x_2,y_2)$ is the Wigner function at $\tau =0$,
and the Green functions $G_\tau(x_j|x^{\prime}_j)$ are given by 
\begin{eqnarray}\fl
G_\tau(x_j|x^{\prime}_j)=\frac{1}{\sqrt{2\pi D^2}}\exp\left[-\frac{
(x_j-x^{\prime}_je^{-\frac12 \gamma \tau})^2} {2 D^2}\right]
\;,\quad D^2=\frac{1}{4\gamma }(1-e^{-\gamma \tau}) \;.
\label{green}
\end{eqnarray}
Remarkably, the diffusion coefficients $D^2$ 
remains positive for all times. Eq. (\ref{fp}) admits a 
stationary solution, which can be easily derived from 
Eq. (\ref{fp}) and, independently on the initial state, 
it has the Gaussian form 
\begin{eqnarray}\fl 
W_{\hbox{\scriptsize stat}}=
\frac{2}{\pi(2M+1)}  
\exp\left(-2\:\frac{x_1^2+y_1^2}{2M+1}\right)\;\times \;
\frac{2}{\pi(2M+1)} 
\exp\left(-2\:\frac{x_2^2+y_2^2}{2M+1}\right) 
\;,  \label{staz}
\end{eqnarray}
corresponding to the (factorized) two-mode thermal density 
matrix given by 
\begin{eqnarray}
\varrho_{stat} = \mu_M \otimes \mu_M\:, \label{rhostaz}
\end{eqnarray}
where $\mu_M$ is the density matrix of a thermal state with $M$
thermal photons 
\begin{eqnarray}
\mu_M= \frac{1}{1+M} \left(\frac{M}{1+M}
\right)^{a^\dag a} \;.  \label{thermal}
\end{eqnarray}
Eq. (\ref{thermal}) states that after a long interaction time 
entanglement is totally destroyed, independently on the amount of entanglement 
initially impinged into the channel. This is not unexpected since the losses 
and the amplification noise resulting from the propagation in the 
fibers induce decoherence on the (initially entangled) beams. 
The questions we want to answer are the following: given an initial twin-beam 
$|X\rangle\rangle$ how entanglement is degraded during the propagation ? And how 
long does the entanglement survive ? In order to answer to these questions, we need 
to solve Eq. (\ref{fp}) at a generic time $\tau$. 
The Wigner function $W_\tau(x_1,y_1;x_2,y_2)$ can be obtained 
by the convolution (\ref{conv}), which can be easily evaluated since the initial 
Wigner function of the twin-beam is Gaussian. We have 
\begin{equation}\fl
W_\tau=\left(2\pi \Sigma_+^2 \: 
2\pi \Sigma_-^2\right)^{-1}\: \exp\left[ -\frac{(x_1+x_2)^2}{4\Sigma_+^2}
-\frac{(y_1+y_2)^2}{4\Sigma_-^2} -\frac{(x_1-x_2)^2}{4\Sigma_-^2}
-\frac{(y_1-y_2)^2}{4\Sigma_+^2} \right]\:, 
\label{evolved}\end{equation}
with the variances given by  
\begin{eqnarray}
\Sigma_+^2 = e^{-\gamma\tau} \sigma_+^2+D^2 \qquad 
\Sigma_-^2 = e^{-\gamma\tau} \sigma_-^2+D^2 \label{evolvedVar}\:.
\end{eqnarray}
A quantum state of a bipartite system is {\em separable} if its density
operator can be written as $\varrho=\sum_k p_k \sigma_k \otimes \tau_k$, where
$\{p_k\}$ is a probability distribution and $\tau$'s and $\sigma$'s are
single-system density matrices. If a state is separable the correlations
between the two systems are of purely classical origin. A quantum state which
is not separable contains quantum correlations {\em i.e.} it is 
entangled. A necessary condition for separability is the positivity of the
density matrix $\varrho^T$, obtained by partial transposition of the original
density matrix (PPT condition) \cite{peres}. In general, PPT has been proved 
to be only a necessary condition for separability. However, for some 
specific sets of states PPT is also a sufficient condition. These include 
states of $2\times 2$
and $2 \times 3$ Hilbert spaces \cite{2x}, and Gaussian states (states with a Gaussian
Wigner function) of a bipartite continuous variable system, {\em e.g.} the
states of a two-mode radiation field \cite{geza,simon}.  Our analysis is
based on this results. In fact, the Wigner function of a twin-beam
produced by a parametric source is Gaussian and the evolution inside 
active fibers preserves such Gaussian character. Therefore, we are able to
characterize the entanglement at any time and to give conditions on the fiber
parameters to preserve entanglement after a given fiber length.  The PPT
condition on the density matrix can be rephrased as a condition on the
covariance matrix of the Wigner function $W(x_1,y_1;x_2,y_2)$ of the two
modes. We have that a state is separable iff 
$$ V + \frac{i}{4} \Omega \geq 0 \qquad
\Omega=\left(\begin{array}{cc}J & 0\\ 0 & -J\end{array}\right) \qquad 
J=\left(\begin{array}{cc}0 & 1\\ -1 & 0\end{array}\right)\:,$$
where $$V_{pk}=\langle \Delta \xi_p \Delta\xi_k\rangle=\int d^4 \xi\: 
\Delta \xi_p \Delta\xi_k\: W(\boldsymbol\xi)\:,$$ with
$\Delta\xi_j=\xi_j-\langle\xi_j\rangle$ and
$\boldsymbol\xi=\{x_1,y_1,x_2,y_2\}$. 
\par
Given a Wigner function of the form (\ref{evolved}) we 
have $\<x_j\>=\<y_j\>=\<\Delta x_j \Delta y_k\>=0, \: \forall j,k$ and  
$$\<\Delta x_j^2\>=\<\Delta y_j^2\>=1/2 (\Sigma_+^2+\Sigma_-^2) 
\qquad \<\Delta x_1 \Delta x_2\>=\<\Delta y_1 \Delta
y_2\>=1/2(\Sigma_+^2-\Sigma_-^2)\:,$$ 
such that 
$$
V= \frac 12\left( 
\begin{tabular}{cccc}
$\Sigma_+^2+\Sigma_-^2$ & 0 & $\Sigma_+^2 - \Sigma_-^2$ & 0 \\
0 & $\Sigma_+^2+\Sigma_-^2$ & 0 & $\Sigma_+^2 - \Sigma_-^2$ \\
$\Sigma_+^2-\Sigma_-^2$ & 0 & $\Sigma_+^2 + \Sigma_-^2$ & 0 \\
0 & $\Sigma_+^2-\Sigma_-^2$ & 0 & $\Sigma_+^2 + \Sigma_-^2$ 
\end{tabular}\right)\:. $$
In order to check separability, we diagonalize $V + i/4 \Omega$ and 
impose that all its eigenvalues are greater or equal to zero. It turns 
out that the state described by $W_\tau$ is separable when 
{\em both} the variances satisfies the
condition 
$$\Sigma_+^2 \geq \frac14 \qquad \Sigma_-^2 \geq \frac14\:.$$
\par
\begin{figure}[ht]
\psfig{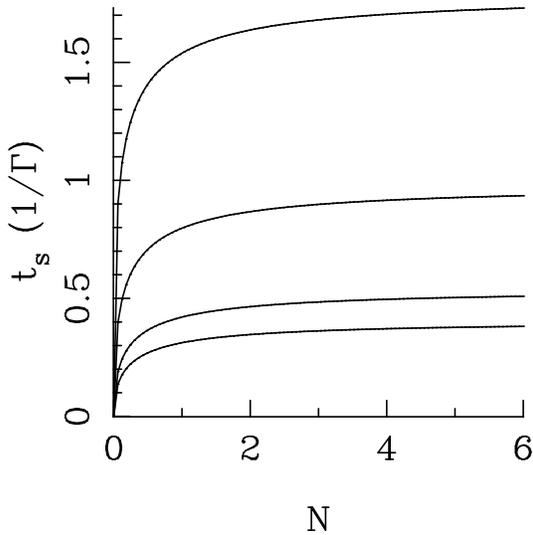}
\caption{The threshold value $t_s$ (in unit $1/\Gamma$) 
as a function of the mean photon number 
of the twin-beam $N$ for different values of the thermal number of photons 
$M$ in the channel. From top to bottom we have the curves for
$M=0.1,0.3,0.7,1.0$.}\label{f:beh}
\end{figure}

The condition $\Sigma_+^2 \geq \frac 14$ is weaker, 
and thus separability is determined by the condition
$\Sigma_-^2 \geq \frac 14 $, which read as follows
$$ \exp\left(-\gamma\tau - 2\lambda \right)+ \frac1{\gamma}
\left[1-\exp\left(-\gamma\tau\right)\right]\geq 1\:. $$ 
Given the parameters $M$, $\Gamma$ and $\lambda$ the threshold 
value $\tau_s$ above which the state become separable is given by
\begin{eqnarray}
\tau_s &=& -\frac1{\gamma}\: \log \left(\frac{1-\gamma}{1-\gamma e^{-2\lambda}}
\right)= (2M+1)\log \left(1+ \frac{1-e^{-2\lambda}}{2M}\right) 
\:.\end{eqnarray}
In terms of the unrescaled time $t$ the threshold for separability 
reads as 
\begin{eqnarray}\fl t_s=
\frac1{\Gamma}\: \log \left(1+ \frac{1-e^{-2\lambda}}{2M}\right)
= \frac1{\Gamma} \log\left(1+\frac{\sqrt{N(N+2)}-N}{2M}\right)
\label{thre}\;.
\end{eqnarray}
In Fig. \ref{f:beh} we report $t_s$ (in unit $1/\Gamma$) 
as a function of $N$ for different values
of the thermal number of photons $M$ in the channel. The threshold $t_s$ 
increases with $N$ from zero to a saturation value given by 
$\log[1+(2M)^{-1}]/\Gamma$. This means that low thermal noise, and 
for an excited entangled state at the input, the entanglement is present 
on a long time scale comparable with the photon lifetime in the medium. The
corresponding threshold $L_s=ct_s$ on the interaction length is of the order 
of many Kilometers assuming a small value for $M$ (which is a reasonable 
approximation at room temperature and optical frequencies), and an attenuation 
factor about $0.3\: dB/Km$ for the optical fibers. \par
A question arises whether or not twin-beam after evolution can be used for 
entanglement based protocols such teleportation. If $\sigma$ is the input 
state of a teleporting machine the Wigner function  
of the output state $\varrho$ (the teleported state) is given by 
$$ W[\varrho](x,y) =\!\!\int\!\!\!\!\int\!\!\!\!\int\!\!\!\!\int\!\!  
dx_1 dy_1  dx_2 dy_2 \: W[\lambda] (x +x_2,y +y_2, x_1+x_2,-y_1-y_2)\:
W[\sigma] (x_2,y_2)\:,$$
where $W[\sigma] (x,y)$ is the Wigner function of the input and $W[\lambda]$
that of the entangled state providing the nonlocality for teleportation. For 
pure state at the input $\sigma=|\psi\rangle\langle\psi|$, 
the teleportation fidelity is given by $F=\langle\psi|\varrho |\psi\rangle$
and can be calculated as the overlap of the Wigner functions $F= \pi \int dx dy 
\: W[\sigma](x,y)\: W[\varrho](x,y)$. For the teleportation of a coherent state 
$|\psi\rangle=|z\rangle$ supported by evoluted twin-beam of the 
form (\ref{evolved}) the fidelity is given by $$ F = \frac{1}{1+ 4 
\Sigma_-^2}= \frac{1}{1+e^{-2\lambda-\Gamma t}+(1-e^{-\Gamma t})(2M+1)+
(1-\eta)/\eta} \:.$$ 
In the ideal case of entanglement provided by unperturbed 
twin-beam we have $\Sigma_-=\sigma_-$, and thus the known formula 
$F=(1+e^{-2\lambda})^{-1}$. Teleportation of a coherent state is a truly quantum 
(nonlocal) protocol if the fidelity is larger than one half \cite{fur}.
Remarkably, the condition $F\geq1/2$ is equivalent to the condition 
for entanglement survival $\Sigma_-^2 \geq \frac14$. This means that evolved
twin-beam can be used for teleportation in the whole range of parameters
preserving entanglement, {\em i.e.} that no regimes of {\em unusable} entanglement 
are present. \par
The evolution and the degradation of continuous variable entanglement of 
twin-beam in optical fibers have been studied by means of two-mode Wigner 
function and PPT condition for separability. 
During the propagation, entanglement survives for a finite interval of time.
In terms of the damping rate $\Gamma$ and the thermal noise $M$, the threshold 
value for separability 
ranges from $t_s=0$ to $\log[1+(2M)^{-1}]/\Gamma$, increasing with the initial
squeezing parameter $\lambda$. For low thermal noise, and for an excited entangled 
state at the input, the entanglement is present 
on a long time scale compared with the photon lifetime in the medium. 
Evolved twin-beam supports coherent state quantum teleportation in  
the whole range of parameters preserving entanglement.
\section*{Acknowledgments}
This work has been supported by INFM through project PRA-CLON-2002
and by EEC trough project TMR-2000-29681 (ATESIT).

\end{document}